\documentclass[11pt]{amsart}
\usepackage[margin=1in]{geometry}
\usepackage{amssymb}
\usepackage{booktabs}
\usepackage{xcolor}
\usepackage{enumitem}
\usepackage[colorlinks=true,linkcolor=blue!50!black,citecolor=blue!50!black,urlcolor=blue!50!black]{hyperref}
\usepackage[capitalise,nameinlink]{cleveref}
\usepackage{graphicx}

\newtheorem{theorem}{Theorem}[section]
\newtheorem{lemma}[theorem]{Lemma}
\newtheorem{proposition}[theorem]{Proposition}
\newtheorem{corollary}[theorem]{Corollary}
\theoremstyle{definition}

\theoremstyle{remark}
\newtheorem{remark}[theorem]{Remark}

\newcommand{\df}{!!}

\numberwithin{equation}{section}

\begin{document}

\title[One-component  networks in closed form]{Counting  doubly labelled trees  and  one-component networks}

\author{Louxin Zhang}
\address{Department of Mathematics, National University of Singapore, Singapore 119076}
\email{matzlx@nus.edu.sg}

\date{\today}


\begin{abstract}
A phylogenetic network is one-component if the child of every reticulation is a leaf. This class of networks has been widely used as building blocks in the combinatorial study of phylogenetic networks. We establish a bijection between one-component networks on $k$ taxa and rooted binary multi-labelled trees in which every label occurs exactly twice. Using this bijection and Burnside's lemma, we derive a closed formula and a two-term recurrence for the number of one-component networks. We further obtain explicit formulas for counting one-component networks with a prescribed number of reticulations.
\end{abstract}

\maketitle

\section{Introduction}

Phylogenetic networks extend phylogenetic trees by allowing reticulate evolutionary events such as hybridisation, recombination, and horizontal gene transfer \cite{HRS10,Steel16}. To develop efficient algorithms for inferring phylogenetic networks from DNA sequences and gene trees, various classes of networks defined by combinatorial constraints have been introduced. Among these, one-component networks are those in which the child of every reticulation is a leaf. One-component networks coincide with one-component galled trees \cite{GRZ20} and are also called simplicial networks \cite{SteelFuchs}.

This class of networks has been widely used as building blocks in the study of phylogenetic networks. For example, a galled network \cite{HK07} can be decomposed into one-component networks arranged in a tree-like fashion, while a reticulation-visible network can be decomposed into one-component networks connected through a tree-child backbone \cite{GYZ19}. Consequently, counting one-component networks is a fundamental step towards enumerating galled networks \cite{FYZ21,GRZ20} and even unrestricted phylogenetic networks \cite{YZ26}.

Gunawan, Rathin and Zhang \cite{GRZ20} established a connection between one-component networks and rooted binary trees in which each label occurs at most twice (DUP-trees) and no twin-cherry occurs. They derived a higher-order recurrence for counting DUP-trees and, consequently, one-component networks. Fuchs, Yu and Zhang \cite{FYZ21} later determined the first-order asymptotics for the number of galled networks. Chang and Fuchs \cite{CF24} obtained exact formulas for counting galled networks with at most three reticulations. They also showed that, for a fixed number $i$ of reticulations, the number $O_{k, i}$ of one-component networks with $k$ leaves is $(2(k-i)-3)\df$ times a polynomial in $k$ of degree $2i$, but did not determine this polynomial explicitly. To date, no general closed-form formula for $O_{k, i}$ has been obtained.

In this paper, we exploit a bijection between one-component  networks on $k$ taxa and fully doubled DUP-trees, in which every label occurs exactly twice, to derive a closed formula for the number $O_k$ of one-component networks on $k$ leaves. We also show that $O_k$ satisfies a two-term recurrence. Combined with an inclusion--exclusion argument, these results yield explicit formulas for the number $O_{k, i}$ of one-component networks on $k$ leaves with $i$ reticulations.

\section{Preliminaries}

\subsection{Binary phylogenetic trees and  networks}
A \emph{ binary phylogenetic network} (BPN) on a set $X$ of taxa is a rooted acyclic digraph without parallel edges such that 
\begin{itemize}
\item its root has indegree 0 and outdegree 1, 
\item its leaves have indegree one and outdegree 0, and are bijectively labelled by $X$, and
\item its other nodes are either \emph{tree nodes} having indegree one and outdegree two or \emph{reticulations} having indegree two and outdegree one. 
\end{itemize}

\begin{figure}[t!]
\includegraphics[width=0.9\textwidth]{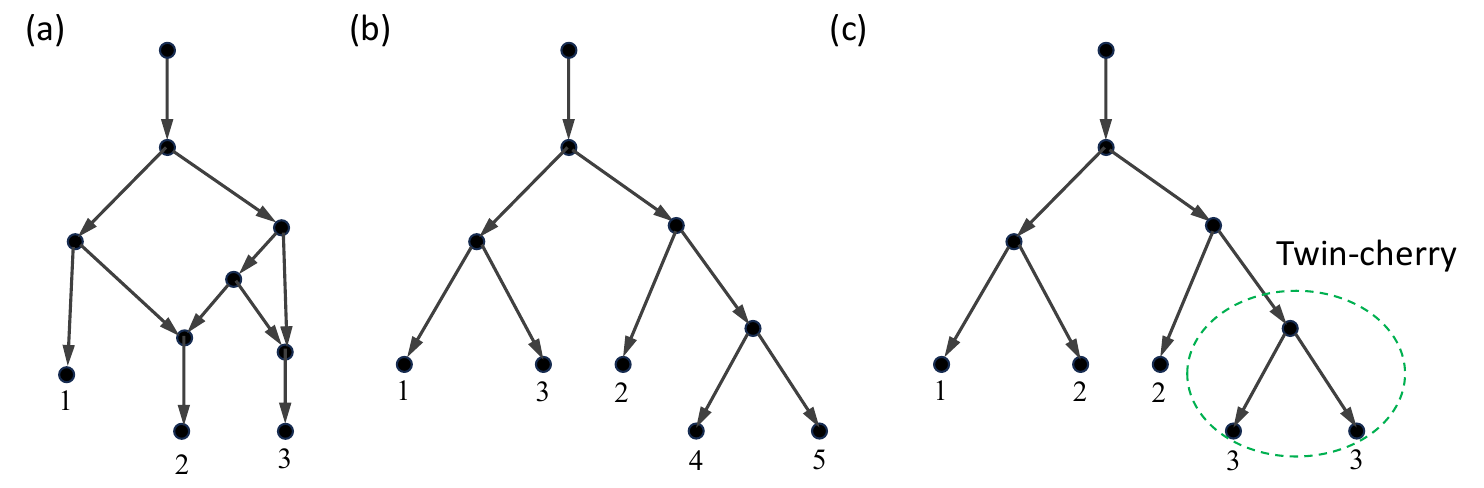}
\caption{Network examples. (a) A one-component network on $\{1, 2, 3\}$. (b) 
A phylogenetic tree on $\{1, 2, 3, 4, 5\}$. (c) A DUP-tree, in which two leaves with label 3 form a twin-cherry.
\label{fig1}}
\end{figure}

Let $u$ and $v$ be two nodes of a BPN. Node $u$ is the {\it parent} of node $v$ and $v$ is a {\it child} of $u$ if 
$(u, v)$ is a directed edge of the network. 
%
A BPN is \emph{one-component} if the child of each reticulation is a leaf (Figure~\ref{fig1}a).  A component of a BPN is a maximal connected subgraph induced by its non-reticulation nodes. It is not hard to see that a BPN is a one-component network if and only if the parents of every reticulation belong to the component containing the root of the network \cite{GRZ20}. 

Let $O_k$ denote the number of one-component  networks on $[k]=\{1,\dots,k\}$, where $k\geq 1$, and $O_{k, i}$ the number of those with exactly $i$ reticulations. Clearly, $O_{k, 0}$ is the number of binary phylogenetic trees on $[k]$, which is 
$\frac{(2k-2)!}{2^{k-1}(k-1)!}$.


A  binary phylogenetic tree is a BPN with no reticulations (Figure~\ref{fig1}b). 
It has $2k-1$ edges if it contains $k$ taxa.
A binary phylogenetic tree on a leaf set $[k]$ is determined by its set of \emph{clusters}, where each cluster is the set of leaves descended from a node. 
The set \([k]\) and every singleton \(\{x\}\), for \(x\in[k]\), are clusters of every such tree. Any two clusters of a tree are either nested or disjoint.


\subsection{Multi-labelled trees}
A \emph{multi-labelled tree} (MUL-tree) is a binary phylogenetic tree whose leaves carry labels, with repetitions allowed, considered up to label-preserving isomorphism 
\cite{HM_JMB06}.
A DUP-tree is a MUL-tree in which  each label is assigned to at most two leaves
 (Figure~\ref{fig1}).
A \emph{twin-cherry} is a pair of sibling leaves with the same label. 

It is clear that a DUP-tree with $s$ singleton labels and $i$ labels of multiplicity two contains $s+2i$ leaves.

For nonnegative integers \(s\) and \(i\), let \(M(s,i)\) denote the set of DUP-trees on the label set \([s+i]\), in which the first \(s\) labels occur once and the remaining \(i\) labels occur twice, and define
\begin{eqnarray}
T(s,i)=\{\text{ DUP-trees in \(M(s,i)\)  without twin-cherries}\}.
\end{eqnarray}
Also define
\begin{equation}
m_{s, i}=|M(s, i)|, \qquad t_{s, i}=|T(s, i)|
\end{equation}
with the convention $m_{0, 0}=t_{0, 0}=0$.

\begin{lemma}\label{lem:sieve}
For all $s,i\ge0$,
\[
m_{s, i}=\sum_{l=0}^{i}\binom{i}{l}t_{s+l,i-l},\qquad
t_{s,i}=\sum_{l=0}^{i}(-1)^{l}\binom{i}{l}m_{s+l,i-l}.
\]
\end{lemma}

\begin{proof}
Given a DUP-tree with $s$ singleton labels and $i$ doubled labels, let $\Lambda$ be the set of doubled labels whose two leaves form a twin-cherry. Contracting each of these twin-cherries to a single leaf gives a DUP-tree with $s+|\Lambda|$ singleton and $i-|\Lambda|$ doubled labels. It has no twin-cherry: a new one would need a third leaf with the same label, but no label occurs more than twice. Conversely, from $\Lambda$ and such a twin-cherry-free tree, replacing each leaf with label in $\Lambda$ by a twin-cherry recovers the original. This proves the first identity, and binomial inversion gives the second.
\end{proof}

\subsection{Double and rising factorials}
 We use the conventions $(-1)\df=1$ and $(-3)\df=-1$, so that $(2m-3)\df=(2m-3)\,(2m-5)\df$ holds for every $m\ge 1$.

 The rising factorial is defined by
\[
(a)_n=a(a+1)(a+2)\cdots(a+n-1),
\]
where \(n\geq 1\), with \((a)_0=1\).

\section{One-component networks and fully doubled DUP-trees}

A DUP-tree  is said to be \emph{fully doubled} if every label appears exactly twice.

\begin{theorem}\label{thm:bij}
For every $k\ge1$, one-component networks on $[k]$ are in bijection with fully doubled DUP-trees on $[k]$. A network with $i$ reticulations corresponds to a tree with exactly $k-i$ twin-cherries. Consequently,
\[
O_k= |M(0,k)|=m_{0, k},\qquad O_{k, i}=\binom{k}{i}\,
t_{k-i,i}.
\]
\end{theorem}

\begin{proof}
We first describe a map from  one-component networks to DUP-trees.  Consider a one-component network $N$. Since every
reticulation has a leaf child in $N$, no reticulation can be a parent of another
reticulation.  Replace each reticulation with leaf child labelled $x$ in $N$ by
two leaves labelled $x$, one attached to each of its two parents.  The
parents are distinct, because parallel arcs are forbidden.  This
operation preserves the two outgoing incidences of every tree vertex
and eliminates all reticulations.  Next replace each original leaf
labelled $y$ (that is, a leaf not below a reticulation) by a cherry
whose two leaves both have label $y$.  The result is a rooted binary
DUP-tree $D(N)$ with two occurrences of every label.

We verify the assertion about twin-cherries.  The two copies of a
reticulation-child label have different parents, so cannot form a
cherry.  Conversely, each original leaf produces exactly one
same-labelled cherry.  Because each label appears only twice, there
are no other twin-cherries.  Thus $D(N)$ has precisely $k-i$
twin-cherries if $N$ has $i$ reticulations.

For the inverse, take a fully doubled DUP-tree $U$.  For each
 twin-cherry,  replace the entire rooted cherry
subtree by a single leaf; this also covers a cherry at the root.
No new twin-cherry can arise, because each contracted label now
occurs once, while the two occurrences of any uncontracted label
were not siblings before contraction.  For every remaining doubled
label $x$, with distinct leaf parents $u$ and $v$, delete both
$x$-leaves and introduce a reticulation $r_x$ with arcs
$u\to r_x$, $v\to r_x$, and $r_x\to x$, where the last $x$  denotes a newly introduced leaf labelled \(x\).  Distinctness of $u,v$ rules out parallel arcs.
The original internal tree vertices and the arcs between them form
one connected rooted tree: removing leaves does not disconnect the
internal-vertex skeleton of a rooted binary tree.  All newly added
arcs point from that skeleton to reticulations and then to leaves,
so no directed cycle is created.  Every internal tree vertex retains
outdegree two, and each reticulation has indegree two and outdegree
one.  Hence the result is a one-component  network.

These operations are inverse on label-preserving isomorphism classes:
contracting the introduced cherries reverses the doubling of original
leaves, and folding each remaining pair reverses the splitting of a
reticulation.  Finally, choose the $i$ labels which do not form
twin-cherries, and contract the other $k-i$ cherries.  The contracted
tree has $k-i$ singleton labels and $i$ doubled labels, without a
twin-cherry; conversely it uniquely expands back.  This leads to
the two identities in the theorem.
\end{proof}

\begin{remark}
The point of \cref{thm:bij} is that the total count needs no twin-cherry-free condition at all, which is introduced in \cite[Theorem~4.2]{GRZ20}. This is what makes a closed formula possible.
\end{remark}

\section{Counting DUP-trees}
\subsection{Trees invariant under swapping pairs of leaves}

Fix a set $S$ of $t$ \emph{single} leaves and $j$ \emph{pairs} $\{x'_b,x''_b\}$, $1\le b\le j$, all $t+2j$ leaves being distinct. Let $\sigma$ be the involution of $S\cup\{x'_b,x''_b\}$ that swaps $x'_b$ and $x''_b$ for every $b$ and fixes $S$. A binary phylogenetic tree with these leaves is \emph{$\sigma$-invariant} if $\sigma$ maps its set of clusters onto itself. 
In that case, \(\sigma\) induces an automorphism of the rooted tree and hence acts on its edges.
Let $\Phi(t,j)$ denote the number of $\sigma$-invariant rooted binary trees.

\begin{lemma}\label{lem:Phi}
For all $(t,j)\ne(0,0)$,
\[
\Phi(t,j)=(2t-3)\df\prod_{b=1}^{j}(2t+4b-5).
\]
\end{lemma}

\begin{proof}
For $j=0$, the condition is empty, so $\Phi(t,0)=(2t-3)!!$
for $t\ge 1$. 

For $(t,j)=(0,1)$, the only tree
is the cherry on $\{x'_1,x''_1\}$, which is
invariant, and the formula gives
$(-1)(-1)=1$.

It therefore suffices to prove that
\begin{equation}\label{eq:Phi-rec}
\Phi(t,j)=(2t+4j-5)\Phi(t,j-1)
\end{equation}
whenever $j\ge 1$ and $t+j\ge 2$.

Write $x'=x'_j$ and $x''=x''_j$, and let
$\sigma'$ be the restriction of $\sigma$ to
the remaining $L=t+2j-2\ge 1$ leaves.
Consider a $\sigma'$-invariant rooted binary
tree $\tau'$ on these leaves. Since $\sigma'$
preserves the cluster set of $\tau'$, it
induces an automorphism of $\tau'$ that
fixes the root and acts on its edges.

Under our convention that the root has
outdegree one, $\tau'$ has $2L-1$ edges.
We construct $\sigma$-invariant trees by
inserting $x'$ and $x''$ into $\tau'$,
according to the action of $\sigma'$ on
its edges.

\medskip
\noindent
\textit{Case (a): An edge $e$ is fixed by
$\sigma'$.}
Subdivide $e$ by a new vertex $u$ and
attach to $u$ a cherry with leaves $x'$
and $x''$. Since $e$ is fixed by $\sigma'$
and the new cherry is preserved by swapping
$x'$ and $x''$, the resulting tree is
$\sigma$-invariant.

\medskip
\noindent
\textit{Case (b): Two distinct edges $e$
and $e'$ are interchanged by $\sigma'$.}
Insert $x'$ into $e$ and $x''$ into $e'$
by subdividing each edge and attaching
the corresponding leaf to the new vertex.
Since $\sigma'$ interchanges $e$ and $e'$,
the resulting tree is $\sigma$-invariant.
Interchanging the positions of $x'$ and
$x''$ gives a second, distinct,
$\sigma$-invariant tree.

\medskip
Conversely, let $\tau$ be a
$\sigma$-invariant rooted binary tree.
Delete $x'$ and $x''$ and suppress the
resulting vertices of indegree one and
outdegree one. The resulting tree $\tau'$
is $\sigma'$-invariant.

If $x'$ and $x''$ form a cherry in $\tau$,
their common parent is fixed by the
automorphism induced by $\sigma$.
After deleting this cherry and suppressing
its parent, the position of the cherry
corresponds to an edge of $\tau'$ fixed
by $\sigma'$. Thus, $\tau$ arises from
Case (a). This also covers the case in
which the cherry is attached to the
stem edge.

Otherwise, $x'$ and $x''$ have distinct
parents. Since $\sigma$ interchanges
$x'$ and $x''$, it also interchanges their
parents. After deleting the two leaves
and suppressing their parents, their
positions correspond to two distinct
edges of $\tau'$ interchanged by
$\sigma'$. Hence, $\tau$ arises from
Case (b).

These constructions are mutually inverse.
If $\tau'$ has $f$ edges fixed by $\sigma'$
and $p$ pairs of edges interchanged by
$\sigma'$, then it gives rise to
\[
f+2p=2L-1=2t+4j-5
\]
distinct $\sigma$-invariant trees.

Summing over all $\sigma'$-invariant trees
$\tau'$ proves \eqref{eq:Phi-rec}.
Iterating the recurrence, together with
the initial cases, gives
\[
\Phi(t,j)
=(2t-3)!!\prod_{b=1}^{j}(2t+4b-5).
\]
\end{proof}


\subsection{Explicit counting formulas}

\begin{theorem}\label{thm:M}
For all $s,i\ge0$ with $(s,i)\ne(0,0)$, the number $m_{s, i}$ of DUP-trees with $s$ singleton  labels and $i$ doubled labels is given by
%
\begin{equation}\label{eq:Mclosed} m_{s, i}=\frac{P_s(i)}{2^{i}}\sum_{m=0}^{i}\binom{i}{m}Q_s(m),\end{equation}
where
\begin{equation} \label{eq:Mclosed2} 
P_s(i)=(2s-3)\df\prod_{a=1}^{i}(2s+4a-5),\quad
Q_s(m)=\prod_{a=1}^{m}(2s+4a-3).
\end{equation}
\end{theorem}

\begin{proof}
Distinguish the two occurrences of each doubled label \(x\) by assigning them the names \(x'\) and \(x''\).
The group $\Gamma=(\mathbb{Z}_2)^{i}$ of copy-swaps acts on the $(2(s+2i)-3)!!$ rooted binary trees on these $s+2i$ distinct leaves, and the DUP-trees counted by $m_{s,i}$ are exactly the orbits. By Burnside's lemma \cite{Stanley2012}, $m_{s, i}=2^{-i}\sum_{\gamma\in\Gamma}\mathrm{Fix}(\gamma)$. If $\gamma$ swaps the pairs of a set $J$ of $j$ doubled labels, the $2(i-j)$ leaves of the other doubled labels behave as single leaves, so $\mathrm{Fix}(\gamma)=\Phi(s+2(i-j),j)$. This gives the following identity
\begin{equation*}\label{eq:Mburn}
m_{s, i}=\frac{1}{2^{i}}\sum_{j=0}^{i}\binom{i}{j}\,\Phi\bigl(s+2(i-j),\,j\bigr),
\end{equation*}
For \eqref{eq:Mclosed}, put $m=i-j$. By \cref{lem:Phi}, $\Phi(s+2m,i-m)=(2s+4m-3)\df\prod_{a=m+1}^{i}(2s+4a-5)$. Splitting the double factorial into its two residue classes modulo four,
\[
(2s+4m-3)\df=(2s-3)\df\prod_{a=1}^{m}(2s+4a-5)(2s+4a-3),
\]
which holds for all $s,m\ge0$ under the stated conventions. Therefore $\Phi(s+2m,i-m)=P_s(i)\,Q_s(m)$, and summing with weights $\binom{i}{m}$ gives \eqref{eq:Mclosed}.
\end{proof}

For $s\ge1$ and $i\ge1$, $P_s(i)=(2s-1)\df\prod_{a=2}^{i}(2s+4a-5)$. For $s\ge1$ and $i=0$, \eqref{eq:Mclosed} reduces to $m_{s, 0}=(2s-3)!!$. Using the rising factorial notation, $$P_s(i)=(2s-3)\df\,4^{i}\bigl(\tfrac{s}{2}-\tfrac14\bigr)_i$$ and $$Q_s(m)=4^{m}\bigl(\tfrac{s}{2}+\tfrac14\bigr)_m.$$
This leads to the following corollary.

\begin{corollary}\label{cor:Mrec}
Let
\[
\beta_{s,i} =
\begin{cases}
1, & \text{if } (s,i)=(0,0),\\[2mm]
\dfrac{m_{s,i}}{P_s(i)}, & \text{if } s+i\ge 1.
\end{cases}
\]
Then,

(1) The exponential generating function for $\beta_{s, i}$ is given by
\begin{equation}\label{eq:egf}
\sum_{i\ge0}\beta_{s, i}\frac{x^{i}}{i!}=e^{x/2}\,(1-2x)^{-(2s+1)/4},
\end{equation}
and $$\beta_{s, i}=(2i+s-1)\,\beta_{s, i-1}-(i-1)\,\beta_{s, i-2}$$ with $\beta_{s, 0}=1$ and $\beta_{s, 1}=s+1$. 

(2) $m_{s, i}$ satisfies the following recurrence relation
\begin{equation}\label{eq:Mrec}
m_{s, i}=(4i+2s-5)\Bigl[(2i+s-1)\,m_{s,i-1}-(i-1)(4i+2s-9)\,m_{s,i-2}\Bigr]
\end{equation}
for $i\ge2$, provided $(s,i)\ne(0,2)$.
\end{corollary}

\begin{proof}
(1) Write $c=(2s+1)/4$, so that $Q_s(m)=4^{m}(c)_m$. Then
\[
\sum_{i}\Bigl(\sum_{m}\binom{i}{m}Q_s(m)\Bigr)\frac{x^{i}}{i!}=e^{x}\sum_{m}(c)_m\frac{(4x)^m}{m!}=e^{x}(1-4x)^{-c},
\]
and replacing $x$ by $x/2$ accounts for the factor $2^{-i}$; this is \eqref{eq:egf}. The function $B=e^{x/2}(1-2x)^{-c}$ satisfies $(1-2x)B'=(s+1-x)B$. Comparing coefficients of $x^{i-1}/(i-1)!$ gives the recurrence for $\beta_{s, i}$, and $\beta_{s, 1}=(1+(2s+1))/2=s+1$. 

(2) $P_s(i)=(2s+4i-5)P_s(i-1)$ and $P_s(i-1)=(2s+4i-9)P_s(i-2)$; multiplying the recurrence for $\beta_{s, i}$ by $P_s(i)$ gives \eqref{eq:Mrec}. The excluded boundary case $(s,i)=(0,2)$ is the only one that involves $m{0, 0}=0$ rather than $P_0(0)\beta_0(0)=-1$.
\end{proof}

\begin{corollary}\label{cor:T}
 For all $s,i\ge0$ with $s+i\ge1$, 
the count $t_{s, i}$ is given by
\begin{equation}
\label{xxxx}
t_{s, i}=\sum_{l=0}^{i}(-1)^{l}\binom{i}{l}\frac{P_{s+l}(i-l)}{2^{i-l}}\sum_{m=0}^{i-l}\binom{i-l}{m}Q_{s+l}(m). 
\end{equation}
\end{corollary}

\begin{proof}
By the second identity of \cref{lem:sieve},
\[
t_{s, i}=\sum_{l=0}^i(-1)^l\binom{i}{l}m_{s+l,i-l}.
\]
Every summand has at least one label because $s+i\ge1$.
Substituting \eqref{eq:Mclosed} gives the asserted double sum.
\end{proof}

\section{Formulas for one-component  networks}

\begin{proposition}
\label{yyyyy}
For all integers \(k\ge1\) and \(0\le i\le k\),
the number of one-component  networks on $k$ taxa with $i$ reticulations is given by
\[O_{k, i}=\binom{k}{i} \sum_{l=0}^{i}(-1)^{l}\binom{i}{l}\frac{P_{k-i+l}(i-l)}{2^{i-l}}\sum_{m=0}^{i-l}\binom{i-l}{m}Q_{k-i+l}(m),
\]
where 
\begin{eqnarray*} \label{eq:Mclosed2} 
&P_{k-i+l}(i-l)=(2(k-i+l)-3)\df\prod_{a=1}^{i-l}(2(k-i+l+2a)-5),\\
& Q_{k-i+l}(m)=\prod_{a=1}^{m}(2(k-i+l+2a)-3).
\end{eqnarray*}
\end{proposition}
\begin{proof}
The formula follows from $O_{k, i}=\binom{k}{i}t_{k-i,i}$ using (\ref{xxxx}).
\end{proof}

Proposition~\ref{yyyyy} gives an explicit formula for counting one-component  networks, which is simpler than the high-order recurrence formula in \cite{GRZ20}.
For example,
\[
t_{s, 1}=s\,(2s-1)\df,\qquad t_{s, 2}=(s+1)^2(2s+3)\,(2s-1)\df ,
\]
which yield the formula for the count of one-component networks with one reticulation \cite[Eq. 5]{GRZ20}  and with two reticulations in \cite{CF24}.

\begin{theorem}\label{thm:G1}
Let $O_k$ denote the number of one-component networks on $k$ taxa.

(1) For $k\ge1$,
\begin{equation}\label{eq:G1closed}
O_k=\frac{1}{2^{k}}\prod_{j=1}^{k-1}(4j-1)\;\sum_{m=0}^{k}\binom{k}{m}\prod_{j=1}^{m}(4j-3).
\end{equation}

 (2) $O_1=1$, $O_2=6$ and, for $k\ge3$,
\begin{equation}\label{eq:G1rec}
O_k=(4k-5)\Bigl[(2k-1)\,O_{k-1}-(k-1)(4k-9)\,O_{k-2}\Bigr].
\end{equation}
\end{theorem}

\begin{proof}
(1) The bijection of \cref{thm:bij} gives $O_k=m_{0, k}$.
For $k\ge1$, the factor $P_0(k)$ in \eqref{eq:Mclosed}
simplifies, using $(-3)\df=-1$, to
\[
P_0(k)=(-3)\df\prod_{j=1}^k(4j-5)
=\prod_{j=1}^{k-1}(4j-1).
\]
Also $Q_0(m)=\prod_{j=1}^m(4j-3)$.
Substitution gives \eqref{eq:G1closed}. 

(2) 
By Theorem~\ref{thm:bij}, \(O_k=m_{0,k}\). Thus, the recurrence follows immediately from (\ref{eq:Mrec}) in Corollary 4.3 with \(s=0\) and $i=k$, where \(k\ge3\). The initial values \(O_1=1\) and \(O_2=6\) follow from \eqref{eq:G1closed}.
\end{proof}

Using Theorem~\ref{thm:G1}, we are able to give a simple proof of the following known result in \cite{FYZ21}, which was  obtained by bounding the recurrence of \cite{GRZ20}.

\begin{corollary}\label{cor:asym}
As $k\to\infty$,
\[
O_k\sim\frac{e^{1/4}}{4\sqrt{2}\,\pi}\cdot\frac{8^{k}(k!)^2}{k^{2}}
\sim\frac{\sqrt{2\sqrt e}}{4}\,k^{-1}\Bigl(\frac{8}{e^{2}}\Bigr)^{k}k^{2k}.
\]
\end{corollary}

\begin{proof}
Let
\[
A_k=\prod_{j=1}^{k-1}(4j-1),
\qquad
b_k=\frac{O_k}{A_k}.
\]
By (5.1) and Corollary 4.3,
\[
B(x):=\sum_{k\ge 0}b_k\frac{x^k}{k!}
=e^{x/2}(1-2x)^{-1/4},
\]
where $b_0=1$.
Near its unique singularity on the circle of convergence,
$x=1/2$, we have
\[
B(x)=e^{1/4}(1-2x)^{-1/4}
 \bigl(1+O(1-2x)\bigr).
\]
The transfer theorem for algebraic singularities therefore gives
\[
b_k=k![x^k]B(x)
\sim\frac{e^{1/4}}{\Gamma(1/4)}2^k k!k^{-3/4}.
\]
On the other hand, the product formula for the gamma function and
the gamma-ratio asymptotic yield
\begin{align*}
A_k&=\prod_{j=1}^{k-1}(4j-1)
=4^{k-1}\frac{\Gamma(k-1/4)}{\Gamma(3/4)}
\sim\frac{4^{k-1}}{\Gamma(3/4)}k!k^{-5/4}.
\end{align*}
Thus $O_k=A_kb_k$ is asymptotic to
\[
\frac{e^{1/4}8^k(k!)^2}{4k^2\Gamma(1/4)\Gamma(3/4)}
=\frac{e^{1/4}}{4\sqrt2\,\pi}
 \frac{8^k(k!)^2}{k^2},
\]
where the last equality uses Euler's reflection identity
$\Gamma(z)\Gamma(1-z)=\frac{\pi}{\sin (z\pi)}$.

Finally $k!\sim\sqrt{2\pi k}(k/e)^k$ transforms this into
\[
\frac{e^{1/4}}{2\sqrt2}\,k^{-1}
 \left(\frac8{e^2}\right)^k k^{2k}
=\frac{\sqrt{2\sqrt e}}4\,k^{-1}
 \left(\frac8{e^2}\right)^k k^{2k}.
\]
\end{proof}

\Cref{tab:G1} lists $O_k$ and $O_{k, i}$ for small $k$.


\begin{table}[htbp]
\centering
\caption{Numbers $O_{k,i}$ of one-component networks
on $k$ taxa with exactly $i$ reticulations,
and their totals $O_k=\sum_{i=0}^{k}O_{k,i}$.}
\label{tab:G1}
\scriptsize
\resizebox{\textwidth}{!}{%
\begin{tabular}{r|rrrrrrrr|r}
\toprule
$k$ & $i=0$ & $i=1$ & $i=2$ & $i=3$
    & $i=4$ & $i=5$ & $i=6$ & $i=7$
    & $O_k$ \\
\midrule
1 & 1 & 0 &   &   &   &   &   &  
  & 1 \\
2 & 1 & 2 & 3 &   &   &   &   &  
  & 6 \\
3 & 3 & 18 & 60 & 87 &   &   &   &  
  & 168 \\
4 & 15 & 180 & 1,134 & 3,972 & 6,249
  &   &   &  
  & 11,550 \\
5 & 105 & 2,100 & 21,600 & 134,070 & 485,910
  & 804,585 &   &  
  & 1,448,370 \\
6 & 945 & 28,350 & 433,125 & 4,142,700
  & 25,527,825 & 94,432,590 & 161,685,045 &  
  & 286,250,580 \\
7 & 10,395 & 436,590 & 9,287,460 & 126,137,025
  & 1,159,266,150 & 7,110,362,385
  & 26,659,289,790 & 46,726,507,485
  & 81,791,297,280 \\
\bottomrule
\end{tabular}%
}
\end{table}

\section{Concluding remarks}

The twin-cherry-free condition complicates the recurrence derived in \cite{GRZ20}. In this paper, we obtain explicit counting formulas by exploiting a bijection with fully doubled DUP-trees (Theorem 3.1) and applying Burnside's lemma.

 It would be interesting to know whether the total number of  galled networks on $k$ taxa, which is assembled from the numbers $t_{s, i}$ along rooted trees \cite[Theorem~5.1]{GRZ20}, 
 admits an equally simple counting formula.

\end{document}